\documentclass[prl,twocolumn,superscriptaddress]{revtex4}
\usepackage{graphicx}
\begin{document}
\title{Evidence of Luttinger liquid behavior in one-dimensional dipolar quantum gases}
\author{R. Citro}
\affiliation{Dipartimento di Fisica "E. R. Caianiello" and CNISM, Universit\`a degli Studi di Salerno, Salerno, Italy}
\author{E. Orignac}
\affiliation{Laboratoire de Physique de l'\'Ecole Normale
  Sup\'erieure de Lyon, CNRS-UMR5672, Lyon, France}
\author{S. De Palo}
\affiliation{DEMOCRITOS INFM-CNR and Dipartimento di Fisica Teorica, Universit\`a Trieste, Trieste, Italy }
\author{M.~L. Chiofalo}
\affiliation{Classe di Scienze, INFN and CNISM, Scuola Normale
  Superiore, Pisa, Italy}

\begin{abstract}
The ground state and structure of a one-dimensional Bose gas with
dipolar repulsions is investigated at zero temperature by a
combined Reptation Quantum Monte Carlo (RQMC) and bosonization
approach. A non trivial Luttinger-liquid behavior emerges in a
wide range of intermediate densities, evolving into a
Tonks-Girardeau gas at low density and into a classical
quasi-ordered state at high density.  The density dependence of
the Luttinger exponent is extracted from the numerical data,
providing analytical predictions for observable quantities, such
as the structure factor and the momentum distribution. We discuss
the accessibility of such predictions in current experiments with
ultracold atomic and molecular gases.
\end{abstract}

\maketitle

\smallskip

The realization of Bose-Einstein condensation (BEC) in trapped
ultracold quantum atomic gases~\cite{collection} is at the frontier of
modern atomic and molecular, optical and condensed-matter
physics~\cite{review_stringari}. Especially
fascinating experimental results arise from the possibility of tuning
the atomic interactions. Use of Fano-Feshbach
resonances~\cite{feshbach} to change magnitude and sign of
the s-wave scattering length $a$ characterizing the
contact interactions, has allowed {\it e.g.}
the observation of collapsing Bose condensates~\cite{bosenova_1} and of
the crossover from a BEC to a Bardeen-Cooper-Schrieffer-like
transition~\cite{BCS}.

More recent experiments have demonstrated that the range of the
interactions can also be manipulated. Dipole
interactions with long-range anisotropic character have been
observed in $^{52}$Cr atoms~\cite{Dipole} after exploiting the large
magnetic moments of this atomic species, that is $\mu_d\approx 6\mu_B$
with $\mu_B$ being the Bohr magneton. A  BEC containing up to 50000 $^{52}$Cr atoms
has then been obtained below a transition temperature $T_c\simeq$ 700$nK$~\cite{BECCr} and its dynamical
behavior is being investigated~\cite{BECCrDyn}. Promising proposals to tune and shape
the dipolar interaction strength in
quantum gasees of heteronuclear polar molecules have more recently
been suggested~\cite{zoller}.
Significant theoretical predictions have accompanied such
realizations~\cite{baranov}. The stability diagram
of anisotropic confined dipolar gases has been predicted to be
governed by the trapping geometry~\cite{DipStab,Giova02}, as
corroborated by Path-Integral QMC
studies~\cite{Nho05}. Different conclusions are reached by more recent
Diffusion QMC simulations including the dependence of $a$ on the dipole interaction~\cite{Bohn}.

Tuning of the interactions can be combined with the
enhancement of quantum fluctuations after reducing their
dimensionality by {\it e.g.} storing them in elongated
traps~\cite{Dip_BEC_1D,OL}, which could be relevant to applications such as precision
measurements~\cite{PrecExp}, quantum computing~\cite{QuComp},
atomtronic quantum devices, and theoretical investigations of novel quantum
phase transitions~\cite{NovelQPh}.

In the case of quasi one-dimensional (1D)
condensates with short-range interactions, a rich phenomenology is known to emerge from the
collective character of the single-particle degrees of

freedom, despite the absence of broken symmetries~\cite{giamarchi_book_1d}.
Bosons are known to arrange in a Luttinger-liquid state, with single particles being replaced
by collective density excitations~\cite{haldane_bosons,Lutt_BEC1D}.
Strong repulsion may also lead to the fermionization of
interacting bosons in the so-called
Tonks-Girardeau (TG) regime~\cite{girardeau_bosons1d,schultz_1dbose,TG_real}. Experiments in
elongated traps have provided evidence for such 1D
fluctuations~\cite{Dip_BEC_1D}.

In the case of quasi-1D condensates with dipolar interactions, an
interesting question arises whether the quantum fluctuations
are sufficiently enhanced to drive the BEC in a strong-coupling
regime. More recent Diffusion QMC
simulations~\cite{Lozovik} for a homogeneous 1D
dipolar Bose gas have revealed a crossover behavior with increasing
linear density, from a liquid-like superfluid state to an ordered, normal, state with
particles localized at lattice sites.
For low values of the density, the system is seen to behave as a
Tonks-Girardeau gas~\cite{girardeau_bosons1d}.

A consistent theoretical description of the whole crossover region would be useful
to predict the behavior of observable quantities in the experiment. It is well
known~\cite{haldane_bosons} that the low-energy properties of
a one-dimensional system of interacting bosons can be obtained from an
effective quadratic Hamiltonian within a
Luttinger-liquid theory, provided that the
interaction is sufficiently short ranged. Besides,
1D fermionic models with interactions falling off as $1/x^\alpha$
have been studied in the past by many
authors~\cite{tsukamoto99_longrange1d,inoue06_longrange1d}, concluding
that for $\alpha>1$ the long-range part of the
interaction decays rapidly enough to preserve the linear energy
dispersion $\omega(k)=u |k|$ typical of short-range models.

In this Letter we provide evidence for a robust Luttinger-liquid
behavior of a 1D dipolar Bose gas in the crossover region,
evolving into a Tonks-Girardeau gas at low density and into a
classical quasi-ordered state at high-density. We reach this
conclusion by comparing theoretical results from a bosonization
approach against Reptation Quantum Monte Carlo
simulations~\cite{RMC}. By analyzing the scaling behavior of the
static structure factor expected for a Luttinger liquid, we find
small Luttinger exponents at intermediate-to-high density values,
signalling strong repulsions at short distances. The knowledge of
Luttinger parameters is then exploited to determine observable
properties of the fluid in a parameter range accessible to
experiments. To the best of our knowledge this is the first
demonstration of such strongly correlated liquid behavior in
atomic Bose gases with dipolar interactions~\cite{NoteCS}.

We consider $N$ atoms with mass $M$ and permanent dipoles moments
arranged along a line. For the purposes of the present work,
we assume $a\rightarrow 0$ after {\it e.g.} exploiting a Fano-Feshbach
resonance. In effective Rydberg units $Ry^*={\hbar^2}/({2 M r_0^2})$ the
Hamiltonian is
\begin{equation}
H=\left(-\frac{1}{r_s^2}
\sum \frac{\partial^2}{\partial x^2}+
\frac{1}{r_s^3} \sum_{i<j}\frac{1}{|x_i-x_j|^3} \right)\; ,
\label{eq:HdipoleRy}
\end{equation}
where we have defined the effective Bohr radius $r_0\equiv M C_{dd}/(2
\pi \hbar^2)$ in terms of the interaction strength
$C_{dd}$. One has $C_{dd}=\mu_0\mu_d^2$
and $C_{dd}=d^2/\epsilon_0$ for magnetic and electric dipoles
respectively, where $\mu_d$ and $d$ are the magnetic and electric
dipole moments and $\mu_0$ and $\epsilon_0$ are the vacuum permittivities.
The dimensionless parameter governing the system is $r_s=1/(n  r_0)$,
with $n$ being the linear particle density and $r_s r_0$ the unit
length.

We analyse the low-energy structural properties of the fluid. They can be
accessed in experiments with atomic gases by means of {\it e.g.}
Bragg scattering techniques. Here, we infer them
from QMC numerical simulations of the static structure factor $S(k)$
on a box with size $L$ and periodic boundary conditions.
In terms of the density operator $\hat{n}$ this is
\begin{eqnarray}
  \label{eq:def-struc-fac}
  S(k)=\int_0^L dx e^{-ikx} \langle \hat{n}(x)\hat{n}(0)\rangle .
\end{eqnarray}

An accurate size-scaling 
analysis of $S(k)$ is needed to our purposes. To
this aim we resort to the Reptation Quantum Monte Carlo of
Baroni and Moroni~\cite{RMC}, which in
essence is a path-integral technique at zero
temperature giving direct access to the ground-state wavefunction.
This makes the evaluation of the ground-state properties
conceptually simple and the extraction of the correlation
functions in imaginary time immediate~\cite{2dbose}, with the
additional possibility of determining the excitation spectrum.
We use a trial wave-function made up of a
two-body Jastrow factor $\psi_{trial}({R})=\Pi_{i<j}
\exp[{u(|x_i-x_j|)}]$. Removal of the
divergences in $H\psi_{trial}/\psi_{trial}$ due to the potential
energy in the $x\to 0$ limit, is obtained after imposing the cusp
condition, which yields $u(x\rightarrow 0)=-\sqrt{r_s/ x}$~\cite{NoteCusp}.

\begin{figure}[tb]
\includegraphics[width=85mm]{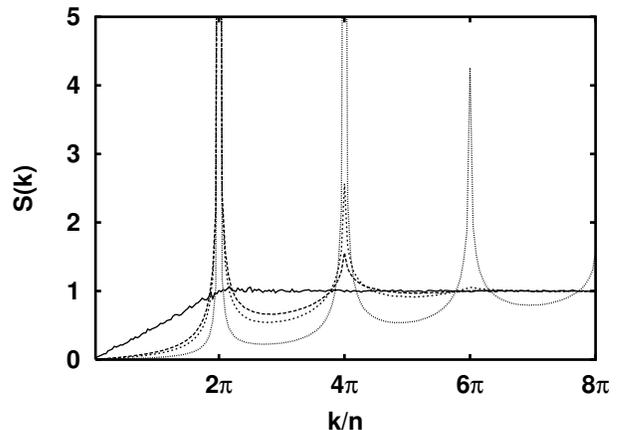}
\vspace{-5mm} \caption{$S(k)$ in dimensionless units
for a dipolar gas with $N=40$
particles and different values of $n r_0=0.01,50, 100$ and
$1000$. Decreasing slopes as $k\to 0$ and the emergence of additional
peaks correspond increasing $nr_0$ values.} \label{fig:sk}
\end{figure}

The RQMC data for $S(k)$ are reported in Fig.~\ref{fig:sk} for $N=40$
and different values of $nr_0=0.01, 50, 100$ and $1000$.
Free fermion-like behavior, typical of the Tonks-Girardeau regime,
is seen at low density values ($nr_0=0.01$ in the
Fig.~\ref{fig:sk}). Quasi-Bragg peaks emerge at reciprocal lattice
vectors $k/n=2\pi m$ ($m$ integer). Their number increases with
increasing $nr_0$, evidencing the occurrence of a quasi-ordered
state. Investigation of the RQMC data at different values of $N$
show that the peak heights scale as $N^{\alpha_m}$ with
$\alpha_m<1$ being an exponent depending on $m$, while their
intensity diverges with the system size.

Both these characteristics are reminiscent of Luttinger-liquid
behavior and therefore we turn on analysing the data within a
bosonization description. In terms of the conjugate bosonic
operators $\phi$ and $\Pi$, the Hamiltonian for a Luttinger liquid
reads~\cite{giamarchi_book_1d}
\begin{eqnarray}
  \label{eq:luttinger-ham}
  H=\hbar \int \frac{dx}{2\pi}\left[ u K (\pi \Pi)^2 + \frac u K (\partial_x
  \phi)^2\right]\; ,
\end{eqnarray}
where $K$ is the Luttinger exponent and $u$ the sound
velocity. Knowledge of $K$ and $u$ allows analytical expressions for
physical quantities of interest.
The structure factor $S(k)$ can be
analytically calculated inserting in Eq.~(\ref{eq:def-struc-fac}) the bosonized expression
of the density operator $\hat{n}$, that is $
  \hat{n}(x)=-\frac 1 \pi \partial_x \phi +  \sum_{m\ne 0} A_m n  e^{2i
  (m\phi(x) -\pi n x)}$~\cite{haldane_bosons,giamarchi_book_1d},
where $A_m$ are non-universal dimensionless constants.
In dimensionless units for $k$, $S(k)$ turns out to be $S_0(k)={K
  k}/({2\pi})$ for $k\to 0$. For $k$ close to $k=2\pi m$ on the scale of the cutoff
  parameter $\pi/\alpha$, we find~\cite{long_paper}:
\begin{eqnarray}
\label{eq:SQ}
  S(k)=\sum_{m\ne 0} A_m^2 n^2 \Sigma_m(k \pm 2 \pi m) \; ,
\end{eqnarray}
where $\Sigma_m(k)$ is explicitly given as
\begin{eqnarray}
   \label{eq:fourier-phi}
   &&\Sigma_m(k)=L(1-e^{-{2\pi\alpha}/{L}})^{2m^2 K}
   \sigma(k;m^2K,L)\\
   &\times& {}_2F_1\left(m^2 K, m^2
   K+{|k|L}/({2\pi});1+{|k|L}/({2\pi});e^{\frac{-4\pi\alpha}L}\right)
   \nonumber\; ,
\end{eqnarray}
where $\sigma(k;m^2K,L)\equiv \Gamma[m^2 K+{|k|L}/({2\pi})][\Gamma(m^2
  K) \Gamma(1+{|k|L}/({2\pi}))]^{-1}$, ${}_2F_1$ the hypergeometric
function.

\begin{figure}[htbp]
  \centering
  \includegraphics[width=85mm]{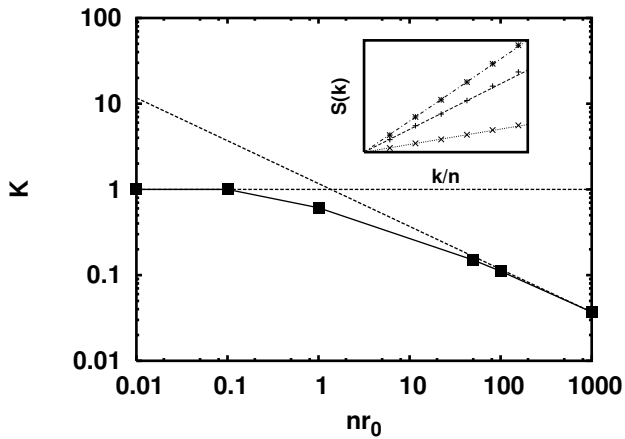}
  \caption{Luttinger exponent $K$ obtained from fitting the RQMC data
  for $S(k)$ in the $k\to 0$ limit. For comparison, the curves
  $K(n\to 0)=1$ and $K(n\rightarrow\infty)= \pi (6\zeta(3)nr_0)^{-1/2}$
  are also displayed as dashed lines, corresponding to the
  Tonks-Girardeau $nr_0\to 0$ limit and to the classical prediction
  for $nr_0\to\infty$. The solid line is a guide to the eye.
  The inset zooms on the low-$k$ portion of $S(k)$ at selected $nr_0$
  (see text).
  }
  \label{fig:luttinger}
\end{figure}

The low-$k$ linear behavior of $S(k)$ can be fitted to extract the
Luttinger $K$ at different densities. The resulting $K(n)$ is
displayed in Fig.~\ref{fig:luttinger}. The inset shows $S(k\to 0)$
for selected densities, where descreasing slopes correspond to
increasing values of $nr_0=50,100,1000$, the symbols representing
the RQMC data and the lines the Luttinger prediction $S(k\to
0)=Kk/(2\pi)$. $K$ remains below the Tonks-Girardeau limit in a
wide range of densities, signalling the occurrence of strong
correlations. As $nr_0<1$, $K$ tends to the Tonks-Girardeau limit
$K=1$. In the opposite high-density limit, $K$ tends to vanish
following the classical-theory prediction
$K(n\rightarrow\infty)\sim \pi (6\zeta(3)nr_0)^{-1/2}$.

$\Sigma_m(k)$ acquires a simplified form depending on whether
$\gamma\equiv m^2K<1/2$ or $\gamma>1/2$ as
$L\to\infty$. For $\gamma <1/2$ we get
$\Sigma_m(k) \sim L^{1-2m^2 K} \sin (\pi
m^2 K)\beta(K;L)$ with $\beta(K;L)\equiv \Gamma(1-2m^2 K) \Gamma(m^2 K + {kL}/({2\pi}))/
  \Gamma(1-m^2 K + {kL}/({2\pi}))$, indeed revealing
the presence in $S(k)$ of quasi-Bragg peaks at $|k|=2\pi m$ and a
scaling behavior in the vicinity of the peak. The peak intensities
diverge with the system size $L$ and their number increases with
decreasing $K$ (namely increasing $n$). The peak's height scales
as $N^{1-2m^2K}$. For $\gamma>1/2$ instead we get $\Sigma_m(k\to
0)\sim  2^{1-2\gamma} [\Gamma(2\gamma -1)][\Gamma(\gamma)^2]^{-1}$
and the peaks disappear.

\begin{figure}[htbp]
  \centering
  \includegraphics[width=85mm]{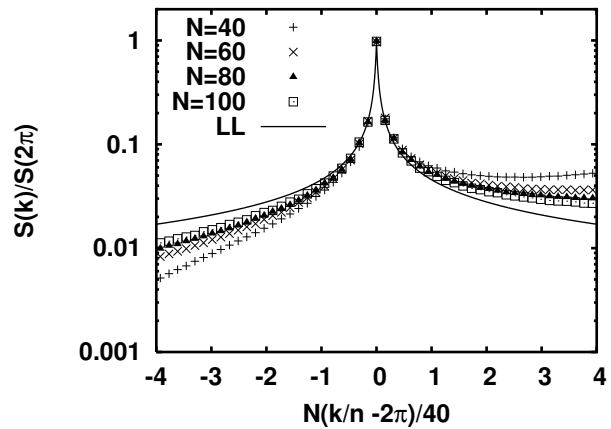}
  \caption{$S(k)/S(2\pi)$ in $n^{-1}$ units {\it vs.} $N(k-2\pi)/40$ at $nr_0=50$ for different $N$
   values (see legend). Symbols represent RQMC data, the solid line is
   the Luttinger-liquid (LL) prediction.}
  \label{fig:SF}
\end{figure}

Besides the linear $k$ behavior,
evidence of a Luttinger liquid comes also from the scaling behavior of $S(k)$. Fig.~\ref{fig:SF}
displays the comparison between the RQMC $S(k)$ close to
$k=2\pi$ towards the quasi-ordered
region with $nr_0=50$, and the predictions of the
Luttinger-liquid theory Eqs.~(\ref{eq:SQ})-(\ref{eq:fourier-phi}) with
$m=1$. We actually plot $S(k)/S(2\pi n)$ for
different $N$ values and verify that they indeed collapse on a single
curve. This signature of Luttinger-liquid behavior against the
RQMC data is robust in the whole crossover region. This is the
central result of the present work. Similar results are
indeed found at the remaining density values above
$nr_0=1$, while the scaling is absent below $nr_0=1$.
The same scaling analysis holds for the second peak at $k=4\pi$ when it is
present. At $k=6\pi$ ($m=3$), we have $2m^2K>1$ for $K<0.055$
and the divergent peak is seen to show up in the simulation
only for the data set with $nr_0=1000$, as expected.

The peak's asymmetry visible in Fig.~\ref{fig:SF} results from
band curvature effects.
Asymmetric structure factors are for example obtained within
exact calculations on Calogero-Sutherland
models~\cite{mucciolo94_cs_dynamical} which indeed possess a
non-linear spectrum, and could be also viewed
as anharmonic phonon effects~\cite{schulz}.

As a consistency check, we have verified that the $K$ values
obtained from the RQMC $S(k)$ agree with those determined from the
RQMC energy per particle $\epsilon$, as follows. Since
Hamiltonian~(\ref{eq:HdipoleRy}) is invariant under Galilean
boosts\cite{haldane_bosons}, we have $u K = {\hbar }\pi n/M$.
Within the bosonization
procedure~\cite{haldane_bosons,giamarchi_book_1d}, the
compressibility is $\chi={K}({\hbar \pi u n^2})^{-1}$. Comparing
this result with the definition of $\chi$ in terms of density
derivatives of $\epsilon$, that is ${\chi}^{-1} =n ^2
(n\epsilon(n))''$, we obtain
$K=\hbar\pi\sqrt{n/[m(n\epsilon(n))'']}$. This result stems from
the trivial fact that the exact RQMC satisfies the compressibility
sum-rule. However, we have also demonstrated on a purely formal
basis that our Luttinger-liquid scheme satisfies the $\chi$
sum-rule too.

The knowledge of $K(n)$ allows us to determine other observables such
as the momentum distribution
$n(k)$, which can be accessed in current atomic BEC's experiments
{\it via} e.g. the analysis of time-of-flight
images. Since in the continuum the boson creation operator is
represented by $\psi^\dagger_B(x) \sim
\exp[{-i\pi \int^x dy \Pi(y)}]$~\cite{giamarchi_book_1d}, we find that
$n(k) \propto k^{1/2K-1}$ as $L\rightarrow \infty$. As the
density increases, the divergence of $n(k)$ at
$k=0$ is expected to be gradually reduced
until it disappears for $K<1/2$.

In conclusion, we have shown that dipolar gases in a reduced 1D
geometry are promising candidates to observe a strong-coupling
Luttinger-liquid behavior with $K<1$. The Luttinger-liquid
behavior is robust over a wide range of $nr_0$ values, evolving
into a Tonks-Girardeau gas at low density and into a classical
quasi-ordered state at high-density. Further testing of the
Luttinger-liquid behavior can be provided by the excitation
spectrum in the crossover~\cite{Giova1D} where, as from
preliminary data, the same combination of bosonization and RQMC
techniques can be successfully applied. Observation of the
Luttinger liquid in the crossover region is within the reach of
current and future experiments in elongated traps. The effective
1D character of trapped dipolar gases is governed by the condition
$n^{-1}\gg l_\perp$, with $l_\perp\equiv
(r_0/(4a_\perp))^{1/5}a_\perp$ ensuring the suppression of forward
scattering~\cite{zoller}. Values of the transverse oscillator
length $a_\perp=\sqrt{\hbar/(M\omega_\perp)}$ related to the
harmonic angular frequency $\omega_\perp$ can be pushed to $50$
nm. Then, dipolar ${\rm SrO}$ molecular gases in elongated traps
are promising candidates, as we might have $l_\perp\approx
0.2\mu\;$m while $r_0\approx 240\; \mu$m, yielding access to
$nr_0$ values up to the quasi-ordered regime. For dipolar ${\rm
{}^{52}Cr}$ gases instead we can have $l_\perp\approx 31$ nm while
$r_0\approx 4.8$ nm, yielding access to $nr_0$ values remaining
down in the Tonks-Girardeau regime. If ultracold ${\rm SrO}$
dipolar quantum gases were to be created in such quasi-1D
conditions, we predict the observation of a power-law behavior of
the peak heights in the structure factor and a flattening of the
momentum distribution with increasing $nr_0$ as fingerprints of
Luttinger-liquid behavior.

\begin{acknowledgments}
We thank Dr. Paolo Pedri for very useful suggestions on
the quasi-1D realizations. RC, SDP and MLC thank Prof. G.~C. La Rocca
for useful discussions. SDP is indebted to Dr. S. Moroni for helpful
discussions and QMC support.
\end{acknowledgments}


\begin{thebibliography}{cccc}
\bibitem{NoteCS} Similar results are known in
lattice models with hard-core~\cite{gomez} or continuum models
with $1/r^2$
repulsions~\cite{calogero_model,mucciolo94_cs_dynamical}.
\bibitem{NoteCusp} We obtain the same results also using
  either the Luttinger form
  $u(x)\propto\sin^\alpha(\pi x/L)$ or gaussians
  centered on the lattice sites as trial wavefunctions.
\end{thebibliography}
\end{document}